\documentstyle[aps,prb]{revtex}
\begin{document}
\twocolumn[]

{\bf Comment on ``Periodic wave functions and number of extended states
in random dimer systems''}

In a recent report,~\cite{lindblad} Huang {\em et al.\/} studied
numerically the electronic properties of a random dimer model (RDM) and
found extended periodic wave functions near the critical energies.  They
also claimed that the number of extended states is proportional to
$\sqrt{N}$, where $N$ is the size of the system, which is consistent
with previous results.

(i) There are {\em no} periodic wave functions in the RDM.  Indeed,
even at the critical energies it was shown previously~\cite{flores}
that the wave functions have a constant norm but the phases are
random.  Essentially the wave function is $\psi_i=e^{i\theta_i}$,
where $\theta_i$ is random.  This does, however, lead to a constant
envelope.  We define the envelope as the ensemble of curves which
include all possible values of the random wave function.  In this
point we present the analytical derivation against the main result
presented in Ref.~\onlinecite{lindblad} and show that the envelope
(and {\em not\/} the wave-function itself) is periodic near the
critical energies as long as the localization length exceeds the size
of the system.

The RDM can be analyzed in terms of the  product of the following two
transfer matrices~\cite{hilke2}
\begin{equation}
T_A=\left(\begin{array}{cc} E-V_A & -1\\ 1 & 0 \end{array}\right)^{2},\quad
T_B=\left(\begin{array}{cc} E-V_B & -1\\ 1 & 0 \end{array}\right)^{2}.
\end{equation}
They reduce to the negative identity matrix when $E=V_A$ or $E=V_B$
(critical energies), which means that $T_A$ and $T_B$ commute.  Close to
the critical energy, i.e., when $\Delta E=E-V_A$ is small, the commutator
$[T_A,T_B]=O(\Delta E)$, as opposed to the case without the dimer
condition, where the commutator is always of order 1.  The total system
is described by a random mixture of the products of the transfer
matrices $T_A$ and $T_B$, thus $T=\prod_{i}T_A^{n_i}T_B^{m_i}$, where
$n_i$ and $m_i$ are random.  Therefore the total product is
$T=T_A^{n}T_B^{m}+(N/4)O(\Delta E)$, where $n=\sum_in_i$, $m=\sum_im_i$
and $N=2(n+m)$.  This clearly demonstrates that the solution is
equivalent to the system with
\begin{equation}
T=(-1)^{n}T_B^{m},
\label{equivalent}
\end{equation}
as long as $N$ is smaller than $1/\Delta E$.

The main result here is that the envelope of Eq.~(\ref{equivalent})
reproduces all the numerical figures of Ref.~\onlinecite{lindblad}.
Indeed Huang {\em et al.\/} studied numerically two cases:

(a) $V_A=-1$, $V_B=1$ and $E=-1+\Delta E$. As the envelope can be obtained
from the solution of $T_B^m$, the corresponding wave-functions can be 
written as  $\psi_i\sim\cos (ki)$, where 
$E-V_B=2\cos (k)$. To first order in $\Delta E$ we obtain  
$k\simeq\pi-\sqrt{\Delta E/2}$.  From this follows trivially
that the general solution has periodic solutions for $2m\sqrt{\Delta
E/2}=\pi p$, where $p$ is an integer.  This leads to $\Delta
E=2\pi^2p^2/(N/2)^2$, where $N/2$ is the number of dimers when they are
equally distributed.

(b) $V_A=-0.5$, $V_B=0.5$ and $E-V_B=-1+\Delta E$. In this case we obtain
$k\simeq2\pi/3-\sqrt{3}\Delta E/2$ and $\Delta E=2\pi p/\sqrt{3}(N/2)$.

In both cases (a) and (b) we obtain the same dependences as in
Ref.~\onlinecite{lindblad}.

(ii) The use of the inverse participation ratio (IPR)
\begin{equation}
\mbox{IPR}=N\sum_{j=1}^N\mid \psi_j\mid^{4},
\end{equation}
where $\psi_j$ is the normalized wave function at site $j$, is very
ambiguous and a complete multifractal analysis is required in
determining the extension of the wave function.\cite{adame} We can
indeed construct delocalized wave-functions with arbitrary IPR.  For
instance, we can consider the following periodic (normalized) wave
function $\psi_j^{2}= T/N \text{ if } j/T = \text{integer}$ and
$\psi_j^{2}=0$ otherwise, where the period $T$ is an integer ($T<N)$
then from (3), we obtain $\mbox{IPR}=T$.  In this way we have a set of
delocalized states with arbitrary IPR.

(iii) The number of extended states is proportional to $\sqrt{N}$,
which means that the relative number of extended states tends to zero as
$1/\sqrt{N}$, therefore the delocalization properties are important only
in small systems.\cite{hilke}

In conclusion, there are {\em no} periodic wave-functions in the RDM but
close to the critical energies there exist periodic envelopes.  These
envelopes are given by the non-disordered properties of the system.

\vskip.5cm

M. Hilke,\footnote{Department of Electrical Engineering,
Princeton University, Princeton, NJ-08544}
J.C. Flores,\footnote{Universidad
de Tarapac\'{a}, Departamento de F\'{\i}sica, Casilla 7-D,
Arica, Chile} and
F. Dom\'{\i}nguez-Adame.\footnote{Departamento de F\'{\i}sica de
Materiales, Universidad Complutense, E-28040 Madrid, Spain}

\vskip.5cm

Part of this work was made possible due to the project FDI-Chile (JCF
and MH).

\end{document}